\documentclass[twocolumn,showpacs,aps,prl]{revtex4}
\usepackage{graphics}
\usepackage{epsfig}
\begin{document}

\title{Atomic vapor-based high efficiency optical detectors with 
photon number resolution} \author{Daniel F. V. James} 
\email{dfvj@t4.lanl.gov} \affiliation{Theoretical Division T-4, 
University of California, Los Alamos National Laboratory, MS B-283, 
P.O. Box 1663, Los Alamos, NM 87545} \author{Paul G. Kwiat} 
\email{kwiat@uiuc.edu} \affiliation{Dept.  of Physics, University of 
Illinois at Urbana-Champaign, Urbana IL 61801-3080} \date{June 7, 
2002} \pacs{42.50.-p, 32.80.-t, 85.60.Gz}

\begin{abstract}
\noindent We propose a novel approach to the important fundamental
problem
of detecting weak optical fields at the few photon level. The
ability to detect with high efficiency ($>$ 99\%), and to
distinguish
the number of photons in a given time interval is a very challenging
technical problem with enormous potential pay-offs in quantum
communications and information processing.  Our proposal diverges from
standard solid-state photo-detector technology by employing an atomic
vapor as the  active medium, prepared in a specific quantum state using
laser radiation. The absorption of a photon will be aided by a dressing
laser, and the presence or absence of an excited atom will be
detected using the ``cycling transition'' approach perfected for
ion traps.  By first incorporating an appropriate upconversion scheme,
our method can be applied to a wide variety of optical wavelengths.
\end{abstract}
\maketitle

Since its introduction more than 100 years ago \cite{Planck},
the notion of light
quanta has remained an elusive and misunderstood concept
\cite{Einstein}.
The ability to detect individual quanta of radiation is of central
importance to fundamental physics.  A direct example of the
ramifications of photon detector efficiencies are the tests of
violations of Bell's inequalities \cite{Clauser}, which reveal
the non-local nature of quantum mechanics.  By rapidly and
randomly measuring the correlations of space-like
separated particles, one can rule out all theories invoking a local
realistic view of nature.  In order for such a test to be
indisputable, however, the detection efficiencies must be very high
\cite{KwiatEPR}, and to date this has not been achieved in any optical
experiment, though an experiment with rapid, random switching has
been performed \cite{Weihs}.
Recently, however, this ``detector loophole'' was closed in an elegant
experiment using entangled ions \cite{Rowe}
(though unfortunately the ions were so close together that the
``locality loophole'' ruled out any unambiguous test of non-locality).
The cold trapped ions were detected with near 100\% efficiency by
the technique of light scattering via a ``cycling''
atomic transition; in brief, each ion is made to emit a large
number of photons, which are then easily detected.

The importance of high efficiency photon detectors is not
confined to fundamental physics.  In addition to
the obvious relevance for metrology, recently there have been a
number of proposals for realizing scalable quantum computing using
only linear optics \cite{KLM}.  A key ingredient
for these proposals is very high efficiency photon-counting detectors
($>99\%$ \cite{Glancy}).
In addition, it is crucial that the detectors be able to distinguish
the {\it number} of incident photons (e.g., tell the difference
between 1 and 2 photons, or more generally, between $n$ and $n+1$).
It has also been noted that such detectors would enable the
preparation of novel quantum states of light, e.g., many-photon
entangled states \cite{Kok}, which could be of great utility in other
quantum information schemes, such as quantum lithography \cite{Boto}.
Finally, there are other applications, such as telecom fiber-based
quantum cryptography, where the low efficiencies and noise levels of
current detectors significantly limit the achievable
distances \cite{Gisin}.

Most modern photon detectors rely to a greater or lesser extent on the 
photoelectric effect: incident photons are converted to individual 
photo-electrons, either ionized into vacuum or excited into the 
conduction band of some semi-conductor.  Either way, one is relying on 
the capability of amplifying single electrons up to detectable levels 
of current in order to produce a tangible signal 
\cite{Superconducting}.  For example, the silicon avalanche 
photodiodes used in many photon counting experiment typically have 
efficiencies $\eta \approx$ 75\% \cite{Kwiat}.  A number of 
experiments have used a variation of this technology, in which the 
silicon is lightly doped.  These ``visible-light photon counters'' 
have displayed $\eta \approx 88\%$, and predicted to be as high as 
95\% \cite{Kwiat,Takeuchi}.  Moreover, they have demonstrated the 
ability to distinguish the number of initial photo-electrons produced 
(which for $\eta \approx 100\%$ is the same as the number of incident 
photons).  Unfortunately, these devices require cooling to 
6K, and display very high dark count rates (up to 50,000 s$^{-1}$), 
undesirable for quantum communication.

We propose a new approach to the problem: instead of converting each 
photon to a single photo-electron, we propose a compound process by 
which a single photon can be converted into many photons.  The basis 
of our proposal is to combine two successful experimental approaches 
of the last few years: the unprecedented ability to coherently slow 
and stop light \cite{Liu}, and the high efficiency scheme for 
projective quantum state measurements in ion traps mentioned above 
\cite{Rowe}.

\begin{figure}[th!]
\center{ \epsfig{figure=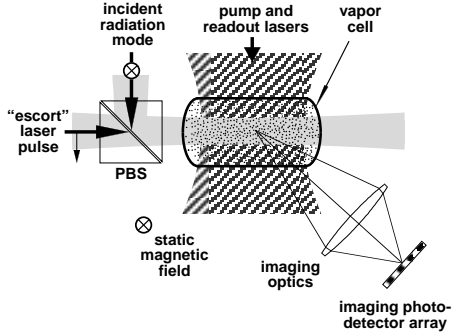,width=60mm}}
\caption{a). Diagram illustrating schematically the proposed photo-detection
technique.  The polarizing beam splitter (PBS) dictates the
polarizations of the applied optical fields, as assumed in deriving the
Hamiltonian eq.(\ref{Ham1}).}
\label{fig1}
\end{figure}

Our scheme consists of a cell containing the vapor of some atomic 
species, e.g., an alkali such as cesium (Fig. 1).  
This vapor will be 
used to coherently absorb the radiation from an incident beam in a 
controlled fashion.  A number of auxiliary lasers prepare the initial 
quantum state of the atoms in the vapor, and control the interaction 
of the atoms with the radiation field.  The radiation to be detected 
is directed into the cell along with an ``escort'' pulse, giving each 
photon some small probability to excite an atom to a metastable state 
-- because there are many atoms, however, the chance that each photon 
is absorbed by one of them can be near unity.  Next a strong read-out 
light is applied, which repeatedly excites any atom in the metastable 
state; the photons resulting from spontaneous decay may then be 
detected.  And because there are many photons, the chance of not 
detecting any at all becomes vanishingly small for realistic detector 
efficiencies.  In fact, if an imaging photon detection scheme is used, 
the number of excited atoms may even be counted, thereby allowing one 
to reliably distinguish input states of different photon number.

Four atomic levels will concern us; $|1\rangle$ and $|2\rangle$ are 
the two sub-levels of the $^{2}S_{1/2}$ ground state (for simplicity 
we will neglect hyperfine structure).  The level $|3\rangle$ is the 
$^{2}P_{1/2}, m_{J}=-1/2$ sublevel, and $|4\rangle$ is the 
$^{2}P_{3/2}, m_{J}=3/2$ sublevel.  The sublevels of the 
$^{2}P_{1/2}$-$^{2}P_{3/2}$ doublet all decay rapidly to the ground 
state, and thus, provided the gas is at a thermal equilibrium 
temperature much less than $T \sim 10^{4} K $, initially there will be 
appreciable population only in the two ground state sublevels 
$|1\rangle$ and $|2\rangle$ (we will see that for other reasons one 
wants to first cool the atoms to $\sim$ 1mK).

The atoms are then prepared in state $|1\rangle$ by optical pumping 
(Fig.2a).  Collisions between atoms will degrade this state 
preparation by exciting population back into state $|2\rangle$.  The 
time taken for such collisions to occur is $\tau_{col} \approx 
\sqrt{M/3 k_{B} T}/n \sigma$, with $M$ the atomic mass, $k_{B}$ 
Boltzmann's constant, $T$ the temperature, $n$ the number-density of 
atoms, and $\sigma$ the collisional cross-section ($\sim 10^{-14} {\rm 
cm}^{2}$ for alkalis).  Collisions can be mitigated in a number 
of ways, such as use of buffer gas, decreasing the atomic density, 
cooling the atoms, using heavier atoms or by raising the energy 
difference between $|1\rangle$ and $|2\rangle$ (thereby making it less 
likely that a single collision will impart sufficient energy to induce 
the transition).  The following two operations (i.e., the photon 
absorption and the readout via a cycling transition) must be 
accomplished in a time considerably smaller than $\tau_{col}$.

\begin{figure}[h]
\center{ \epsfig{figure=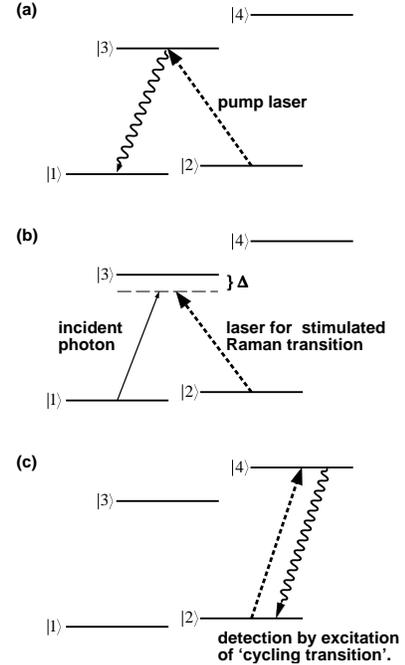,width=60mm}}
\caption{Diagram illustrating the atomic transitions envisioned
for the three-stage photo-detection procedure. Stage (a) is optical
pumping to level $|1\rangle$; stage (b) is the absorption of the
photon with the assistance of a classical ``escort'' pulse; stage (c)
is the detection of any atom in state $|2\rangle$.}
\label{fig2}
\end{figure}
\noindent

Once the atoms are in state $|1\rangle$, the photon field is directed
into the cell, accompanied by a strong ``escort'' laser pulse.  The
resulting two-photon Raman excitation to state $|2\rangle$ is
described by the following interaction picture Hamiltonian:
\begin{eqnarray}
{\hat H}_{I}(t) = \sum^{N}_{i=1} \frac{\hbar \Omega \left({\bf r}_{i},
t \right)}{2} \left(|3\rangle\langle 2|\right)_{i} \exp\left[
i\left(\omega_{32} - \omega_{e}\right) t \right] \nonumber \\
+
i \hbar  \sum^{N}_{i=1} \sum_{\lambda} g_{\lambda, i} {\hat
a}_{\lambda} \left(|3\rangle\langle 1|\right)_{i}
\exp\left[
i\left(\omega_{31} - \omega_{\lambda}\right) t \right] + h.a.
\label{Ham1}
\end{eqnarray}
Here $\Omega \left({\bf r}_{i}, t \right) = \langle 3|{\hat {\bf 
d}}|2\rangle\cdot {\bf{\cal E}}_{e}\left({\bf r}_{i}, t \right) /\hbar 
$ is the Rabi frequency of the $|2\rangle \leftrightarrow |3\rangle$ 
transition of the $i$-th atom due to the escort pulse (represented by 
the analytic signal ${\bf{\cal E}}_{e}\left({\bf r}_{i}, t \right) 
\exp \left(-i\omega_{e} t\right) $, with central frequency 
$\omega_{e}$).  The coefficient $g_{\lambda, i} = 
\sqrt{\omega_{\lambda}/ 2 \varepsilon_{0} \hbar} 
\Phi_{\lambda}\left({\bf r}_{i}\right) \langle 3|{\hat 
d}_{z}|1\rangle$, where $\Phi_{\lambda}\left({\bf r}_{i}\right)$ is 
the spatial mode function for the $\lambda$-th mode of the incident 
field (with annihilation operator ${\hat a}_{\lambda}$ and frequency 
$\omega_{\lambda}$), $\varepsilon_{0}$ is the permittivity of the 
vacuum, and ${\hat {\bf d}}$ is the dipole moment operator.  In 
writing (1), we assume that the escort and photon fields have 
negligible effect on the $|1\rangle \leftrightarrow |3\rangle$ 
transition; this is valid for a particular orientation of static 
magnetic field and specific polarizations for the escort field and the 
photon (see Fig.  1).

We assume that both the photon and the escort field are far detuned
from resonance with the transitions $|1\rangle \leftrightarrow
|3\rangle$ and $|2\rangle \leftrightarrow |3\rangle$, respectively.
In these circumstances, we may adiabatically eliminate the population
of the upper level $|3\rangle$, and the dynamics of the system may be
described to a good approximation by the following effective
Hamiltonian \cite{H-footnote}:
\begin{eqnarray}
{\hat H}_{eff}(t) =
i\hbar  \sum^{N}_{i=1} \sum_{\lambda}
f_{\lambda,i}(t)
&{\hat a}^{\dagger}_{\lambda}& \left(|1\rangle\langle 2|\right)_{i}
\exp\left[
i\left(\omega_{\lambda} - \omega_{0}\right) t \right] \nonumber\\
&+& h.a.
\end{eqnarray}
where $f_{\lambda,i}(t)= \Omega\left({\bf r}_{i},t\right)
g^{\ast}_{\lambda, i}/ 2 \Delta$, $\omega_{0} =
\omega_{21}+\omega_{e}$ and $\Delta = \omega_{31}-\omega_{0} \approx
\omega_{31}-\omega_{\lambda}$.  This is the Hamiltonian describing an
effective two-level system interacting with a quantized field;
the coupling constant $f_{\lambda,i}$ can
be controlled by shaping the profile of the escort pulse.

An important property of this Hamiltonian is that it commutes with the 
``total excitation'' operator $\sum_{i} \left(|2\rangle\langle 
2|\right)_{i} + \sum_{\lambda} {\hat a}^{\dagger}_{\lambda} {\hat 
a}_{\lambda}$.  As a consequence, the number of quanta of the 
radiation field plus the number of excited atoms must be a constant.  
Therefore, once the incident radiation field has been completely 
absorbed, the number of quanta it contained may be determined by 
measuring the number of atoms in the excited state.  If we confine 
ourselves to single photon incident fields, the wavefunction will thus 
have the form
\begin{equation}
|\psi (t)\rangle = \sum_{\lambda} \alpha_{\lambda}(t)
|1_{\lambda}\rangle |g\rangle + \sum_{i} \beta_{i}(t) |vac\rangle
|2_{i}\rangle,
\end{equation}
where $ |g\rangle$ represents the state in which {\em all} of the
atoms are in state  $|1\rangle$; $|2_{i}\rangle \equiv
\left(|2\rangle\right)_{i} |g\rangle $ the state in which all
the atoms are in state $|1\rangle$, except the $i$-th atom which is in
state $|2\rangle$; $|vac\rangle$ is the field vacuum state;
and $|1_{\lambda}\rangle \equiv {\hat
a}^{\dagger}_{\lambda}|vac\rangle$ is the state with one photon in
mode $\lambda$.  The probability amplitudes $\alpha_{\lambda}(t)$
and $\beta_{i}(t)$ obey the equations
\begin{eqnarray}
{\dot \alpha}_{\lambda}(t) &=& -\sum_{i}f_{\lambda, i}(t) \beta_{i}(t)
\exp\left[i\left(\omega_{\lambda}-\omega_{0}\right)t\right],
\nonumber \\
{\dot \beta}_{i}(t) &=& \sum_{\lambda}f^{\ast}_{\lambda, i} (t)
\alpha_{\lambda}(t)
\exp\left[-i\left(\omega_{\lambda}-\omega_{0}\right)t\right] .
\end{eqnarray}
By formally solving for $\alpha_{\lambda}(t) $ , we 
obtain an integro-differential equation for $\beta_{i}(t)$; under the 
Markov approximation and the assumption that we can neglect  
coherence between different atoms (so that superradiant effects are 
negligible), we obtain the following equation for $\beta_{i}(t)$:
\begin{eqnarray}
{\dot \beta}_{i}(t) &=&
\epsilon_{i}(t) \phi_{i}(t) -
\frac{A_{31}}{2} \left|\epsilon_{i}(t)\right|^{2} \beta_{i}(t),
\label{betadot}
\end{eqnarray}
where $A_{31}$ is the spontaneous decay rate of the
$|3\rangle \rightarrow |1\rangle $ transition,
$\epsilon_{i}(t) = \left[i \Omega\left({\bf r}_{i},t \right)
\exp(-i\omega_{0}t) /2 \Delta
\right]^{\ast}$
characterizes the influence of the escort pulse and $\phi_{i}(t) =
\langle 3|{\hat d}_{z}|1\rangle \sum_{\lambda}
\sqrt{\omega_{\lambda}/2 \hbar \varepsilon_{0}} \alpha_{\lambda}(0)
\Phi_{\lambda}\left({\bf r}_{i}\right) \exp\left( -i
\omega_{\lambda}t\right)$ characterizes the action of the
photon we wish to detect.
The first term on the right hand side of Eq.(\ref{betadot}) represents the
absorption of the photon, the second term re-emission. Equation (\ref{betadot})
can be solved in closed form, yielding explicit expressions
(dependent on the photon- and escort-pulse shapes)
for the absorption and scattering probabilities.  In what follows, we
have employed expressions derived from simple square-shaped pulses
for both the photon and the escort pulses (although model dependent
factors of order unity have been suppressed).

Once our single photon is absorbed by one of the atoms, we can detect 
this excited atom with very high probability by employing a third 
auxiliary laser.  This ``readout'' laser is carefully tuned and 
polarized so that any population in each atom's excited state is 
pumped into some convenient upper state $|4\rangle$ (Fig.2c), chosen 
so that it will rapidly decay spontaneously back to the excited state 
{\em only} \cite{Rowe}.  If the readout light persists for some 
microseconds, the atom in the excited state $|2\rangle$ will scatter 
many photons.  By detecting and imaging the scattered radiation, one 
can therefore determine which atom is excited.  The time taken for one 
such photon to be registered is $t_{ro}= \left(2 
\Omega_{r}^{2}+A_{24}^{2} \right)/A_{24}\Omega_{r}^{2} \eta_{det}$, 
$\Omega_{r}$ being the readout laser Rabi frequency, $A_{24}$ the 
decay rate of the $|4\rangle \rightarrow |2\rangle$ transition, and 
$\eta_{det}$ the overall detection efficiency of the imaging system, 
including solid angle acceptance and photon detector efficiency.

The efficiency of the detector we have described is limited by three effects:
the possibility that the photon may be scattered rather than absorbed by
the atom, the possibility that the photon may pass through the medium
without being absorbed by any atom, and the possibility that the atom
which has absorbed the photon may be collisionally de-excited before
the readout can take place.  Using results derived from
eq.(\ref{betadot}) we find that the efficiency is given approximately
by the formula
\begin{equation}
\eta \approx 1- \left( \frac{T_{p} A_{31} \Omega_{e}^2 }{16 \Delta^2 }  \right)^2
-\exp\left[\frac{-q \ell_{cell}}{\ell_{abs}} \right]
-\frac{t_{ro}}{2 \tau_{col}}.
\end{equation}
Here $T_{p}$ is the duration of the photon and escort pulses, 
$\Omega_{e}$ is the (constant) value of the escort pulse Rabi 
frequency, $\ell_{cell}$ is the length of the cell, $q$ is the number 
of passes through the cell made by the photon and escort, and 
$\ell_{abs}=\Delta^2/(\lambda_{ph}^2 n \Omega_e^2 T_{p} A_{13})$ is 
the absorption length of the medium ($\lambda_{ph}$ being the photon 
wavelength).  If we assume a cesium density of $10^{9} {\rm cm}^{-3}$, 
cooled to the Doppler limit of $\sim 10^{-3}$K (thereby rendering 
collisional effects negligible \cite{drift}), a pulse duration $T_{p}$ 
of $10$ nsec, a detuning $\Delta$ of 0.5 GHz, a strong escort pulse 
with $\Omega_e = A_{13}$, a cell length $\ell_{cell} = 2$ mm, a beam 
area $A = 10^{-2}$ mm$^{2}$, $q$ = 100 passes through the cell, and 
$\eta_{det}=1/8$, then the theoretical detection efficiency will be 
$\eta \approx $ 99.8\%.

Multiple photons can also be reliably measured by this technique.  As 
long as the total number of incident photons is much less than the 
total number of atoms participating in the measurement, each photon 
will be absorbed by a different atom, and the number of fluorescing 
atoms observed during readout will be equal to the number of photons.  
Being able to spatially resolve the atoms will assist in counting them 
(the image of the moving atom emitting photons is somewhat reminiscent 
of the tracks in a bubble chamber), though even this is not completely 
necessary, as demonstrated in \cite{Rowe}; as long as the number of 
emitting atoms in each optically resolvable volume is sufficiently 
small that the photon statistics of 1, 2, etc.  atoms radiating are 
distinguishable, the total number of atoms in state $|2\rangle$ can be 
determined.  For example, with $\eta \approx$ 99.8\%, we could in 
principle reliably distinguish states with $\sim$50 photons.

A serious potential problem in realizing this scheme will be ``dark 
counts''.
Provided collisional excitation of state $|2\rangle$ can be kept 
negligible, the principle mechanism by which this can occur is 
excitation out of $|1\rangle$ by the readout laser.  In the example 
discussed above, the readout laser polarized to address the $|2\rangle 
\leftrightarrow |4\rangle$ transition can excite population in 
$|1\rangle$ to the $^{2}P_{3/2}, m_{J}=1/2$ sublevel.  Due to the 
Zeeman splitting, this transition will be detuned by $\delta=2 \mu_{B} 
B / 3 \hbar $ ($\mu_{B}$ being the Bohr magneton and $B$ the magnetic 
field strength), and due to the different dipole moment strengths, the 
Rabi frequency will be reduced by a factor of $\sqrt{3}$.  The 
probability of an atom emitting a dark count photon during readout is 
then $P_{dc} \approx t_{ro} A_{42} \Omega_{r}^{2} /6 
\left(\delta^{2}+\Omega_{r}^{2}/3\right) $.  For a magnetic field $B$ 
= 1 T and $\Omega_{r} = 0.01 A_{42}$, $P_{dc} \sim 2 \times 10^{-5}$.  
The large number of atoms $N$ present in the active medium may make 
dark counts a non-negligible effect (the above parameters yield $N$ = 
20,000, and a net dark count probability of 0.4).  Ways to mitigate 
this problem, e.g., allowing the photon and its escort to pass through 
the atoms multiple times (thereby allowing a reduction in the number 
of atoms without a detrimental effect on the detection efficiency), or 
reducing the area of the optical beams, are a subject of on-going 
investigation.

It may be argued that a detector of this sort is of limited value 
because it operates only for a single frequency, and not, e.g., at the 
telecommunications wavelength of 1550 nm.  If indeed we start out with 
a photon with a different wavelength (but still assumed to be of 
narrow bandwidth), we estimate that by first mixing the photon with a 
strong pulse of an appropriate frequency in a nonlinear optical 
crystal, we can have a near-unity probability of upconverting the 
photon to the frequency needed for our detection scheme.  In this way 
the high efficiency detection method proposed above may be 
used over much of the optical spectrum.

\vspace{0.1 in}
\noindent
{\it Note:  While finishing our manuscript, it came to our
attention that a similar proposal was suggested independently
and essentially simultaneously by A. Imamoglu
(http://xxx.lanl.gov/abs/quant-ph/0205196).}

\noindent
{\em Acknowledgement}: We acknowledge helpful discussions with B. Englert,
P. Milonni, K. Molmer, E. Timmermans, and A. VanDevender. Note: This
work was supported in part by funds from the Advanced Research and
Development Activity and the Los Alamos QUEST initiative.



\begin{thebibliography}{}
\bibitem{H-footnote}
We have neglected A.C. Stark shifts and the Lamb shift --
both may be compensated for by altering the
appropriate polarization mode or laser detuning.
\bibitem{drift}  Low temperatures also ensure that atoms will not move
appreciably during the measurement; the RMS velocity multiplied by the
readout time $t_{ro}$ will be 0.3 mm.
\end{thebibliography}
\end{document}